**Topological dynamics in an optomechanical system with highly non-degenerate modes**

H. Xu,[1] D. Mason,[1] Luyao Jiang,[1] and J. G. E. Harris[1,2]

[1] Department of Physics, Yale University, New Haven, CT, USA, 06511
[2] Department of Applied Physics, Yale University, New Haven, CT, USA, 06511

**Non-Hermitian systems exhibit phenomena that are qualitatively different from those of Hermitian systems[1,2,3,4,5] and have been exploited to achieve a number of ends, including the generation of exceptional points,[6,7,8,9] nonreciprocal dynamics,[10,11] non-orthogonal normal modes,[12,13,14] and topological operations.[9,15] However to date these effects have only been accessible with nearly-degenerate modes (i.e., modes with frequency difference comparable to their linewidth and coupling rate). Here we demonstrate an optomechanical scheme that extends topological control to highly non-degenerate modes of a non-Hermitian system. Specifically, we induce a "virtual exceptional point" between two mechanical modes whose frequencies differ by $>10^3$ times their linewidth and coupling rate, and use adiabatic topological operations to transfer energy between these modes. This scheme can be readily implemented in many physical systems, potentially extending the utility of non-Hermitian dynamics to a much wider range of settings.**

Dissipation is regarded as a disadvantage for many sensing and control applications, and with good reason: dissipation leads to the decay of signals and to the introduction of noise. Nevertheless, there are situations in which precise control over a system may allow dissipation to serve as a resource. For example in a system of coupled harmonic oscillators, damping means that the system is non-Hermitian and so cannot (in general) be described in terms of orthogonal normal modes and real-valued eigenfrequencies. Recent work has demonstrated a number of ways in which non-Hermitian dynamics, non-orthogonal normal modes, and complex-valued eigenfrequencies can be used as a resource. For example, modest changes in the oscillators' parameters can dramatically alter the flow of energy within the system, leading to unidirectional transport and loss-induced transparency,[16,17] chiral modes[18,19], enhanced mode selectivity,[20,21] and loss-induced lasing[22]. In addition, the normal modes' non-orthogonality can lead to interferences in the system's response to its environment[12,13]. These interferences can be used to reduce thermal noise and to achieve improved signal-to-noise ratio for some applications.

Many of these features can be understood as resulting from the presence of an exceptional point (EP) in the system's spectrum. An EP (also known as a non-Hermitian degeneracy) is a degeneracy at which the complex eigenvalues and normal modes both coalesce,[3] and has been observed in a wide range of physical systems.[6,18,23,22,7,8,9,24,19] In the vicinity of an EP, the eigenvalues undergo a generalized version of the avoided crossings that occur in Hermitian systems in the vicinity of a conventional (i.e., Hermitian) degeneracy. The phenomena described in Refs. [16,12,13,14,7,18,19,22] largely reflect the EP's impact upon the spectrum's *local* properties. However the spectrum's *global* properties are also qualitatively altered by the presence of an EP. Specifically, an EP reflects a non-trivial topology in the eigenvalues' dependence upon the system's parameters. In the simplest case, the eigenvalue surfaces possess the same topology as the Riemann sheets of the complex square-root function. This topological structure is most directly manifested in adiabatic transport. Specifically, if a normal mode of the system is initially excited and then the system's parameters are slowly varied in a closed loop (so as to return them to their initial value) that encloses an EP, then the remaining excitation will be transferred to the other normal mode participating in the EP[25,26,27]. On the other hand, if the loop does not enclose the EP then the excitation will remain in the original mode. Additionally, the non-Hermitian dynamics during the operation ensures that this topological energy transfer (TET) is nonreciprocal with respect to both the sense of the control loop and the choice of which mode is initially excited.[5,28,29]

TET and the associated nonreciprocity arise from interplay between the modes' coupling to each other and to their dissipative environment. In practice it is challenging to ensure that both are relevant, and so to date TET has been achieved only between pairs of modes that are nearly degenerate with each other.[9,15] This limits TET to systems in which precise fabrication or *in situ* tuning can be used to realize nearly-degenerate modes. Even then, TET can only occur in a small subset of the mode pairs.

In cavity optomechanical systems, *in situ* tuning can be provided by the interaction of an optical cavity mode with the various vibrational modes of a mechanical element. Elimination of the cavity dynamics leaves the mechanical modes as oscillators whose frequencies, couplings, and dampings can be controlled via laser excitation of the cavity.[30] In previous work on TET,[9] this approach provided the real-time parametric control needed to encircle an EP; however the only accessible EPs were those between the few pairs of mechanical modes whose near-degeneracy was engineered via the mechanical element's precise fabrication.

Here we show that the same optomechanical interaction can be used to eliminate the requirement of near degeneracy. This is accomplished by using an additional laser tone to bridge the frequency gap between well-separated mechanical modes, producing an EP in an appropriately defined frame which we refer to as a virtual EP (VEP). We show that this VEP offers the same features as the conventional one, including non-reciprocal TET.

The system consists of a mechanical oscillator and an optical cavity which are kept at temperature $T = 4.2$ K. The mechanical oscillator is a silicon nitride membrane with dimensions 1 mm × 1 mm × 50 nm. Instead of using a pair of nearly degenerate modes as in Ref. [9], we use two modes whose frequencies are well separated, with bare values (i.e., without light in the cavity) $\omega_1/2\pi = 557.473$ kHz and $\omega_2/2\pi = 705.164$ kHz. The modes' bare damping rates are $\gamma_1/2\pi = 0.39$ Hz and $\gamma_2/2\pi = 0.38$ Hz. When the cavity is driven by a laser the optomechanical interaction alters the mechanical frequencies and linewidths via the well-known optical spring and damping effects.[30] However these effects are not strong enough to couple modes whose frequencies differ by ~ 150 kHz. To overcome this problem, we drive the cavity with two lasers having detunings $\Delta_1 \approx -\omega_1$ and $\Delta_2 \approx -\omega_2$ as illustrated in Fig. 1a. Qualitatively, this ensures that motion of mode 1 (2) at a frequency $\approx \omega_1$ ($\omega_2$) produces a sideband from laser 1 (2) that is approximately resonant with the cavity, and that beating between this sideband and laser 2 (1) causes the intracavity intensity to oscillate at a frequency $\approx \omega_2$ ($\omega_1$). This arrangement (illustrated in Fig. 1b) ensures that near-resonant motion of one mode exerts a near-resonant force on the other mode despite their large frequency difference.

To provide a quantitative description, we consider the equations of motion for the two mechanical modes and the cavity field (including the drive lasers) and then integrate out the cavity field (Methods). The dynamics of the two mechanical modes is then given by the effective Hamiltonian

$$H_{\text{eff}} = \begin{pmatrix} \omega_1 - i\gamma_1/2 + \sigma_{11} & \sigma_{12} e^{i(\Delta_1 - \Delta_2)t} \\ \sigma_{21} e^{-i(\Delta_1 - \Delta_2)t} & \omega_2 - i\gamma_2/2 + \sigma_{22} \end{pmatrix} \qquad (1)$$

where the self-coupling rates

$$\sigma_{nn} = \sum_{k=1}^{2} \frac{-ig_n^2 P_k}{\hbar \Omega_k} \kappa_{in} |\chi(\Delta_k)|^2 (\chi(\omega_n + \Delta_k) - \chi(\omega_n - \Delta_k)) \qquad (2)$$

reflect the usual optical spring and damping from each laser. The coupling between the two modes arises explicitly from the two-laser drive and is given by

$$\sigma_{mn} = \frac{-ig_1 g_2 \sqrt{P_1 P_2}}{\hbar \sqrt{\Omega_1 \Omega_2}} \kappa_{in} \chi(-\Delta_m) \chi(\Delta_n) (\chi(\omega_m + \Delta_m) - \chi(\omega_n - \Delta_n)) \qquad (3)$$

where $m \neq n$. In these expressions $\chi(\omega) = (\kappa/2 - i\omega)^{-1}$ is the cavity susceptibility, $g_{1,2}$ are the optomechanical coupling rates, $P_{1,2}$ and $\Omega_{1,2}$ are the input powers and frequencies of the driving lasers, $\kappa/2\pi = 180$ kHz is the cavity linewidth, and $\kappa_{in}/2\pi = 70$ kHz is the cavity input coupling rate. The system is controlled by varying the powers and detunings of the drive lasers, and the mechanical modes are monitored via a heterodyne measurement that uses a third laser which addresses a separate cavity mode with constant power and detuning (a detailed description of the apparatus is given in Ref. [9]). The single-mode optomechanical effects described by the $\sigma_{nn}$ are orders of magnitude smaller than the separation between the two mechanical modes $(\omega_1 - \omega_2)/2\pi \approx 150$ kHz; however as shown below, the system still possesses an EP that allows for TET.

To demonstrate the existence of an EP we measure the spectrum of the mechanical modes as a function of the laser powers and detunings. We fix $P_1 = P_2 \equiv P$ and define the common detuning $\Delta \equiv \Delta_1 + \omega_1 = \Delta_2 + \omega_2 + \delta$. The parameter $\delta$ determines the overall detuning of the coupling process shown in Fig. 1b (e.g., for $\delta = 0$ the resonant motion of mode 1 (2) drives mode 2 (1) exactly on resonance), and so is chosen to be comparable to the mechanical linewidths. For the measurements shown $\delta/2\pi = 100$ Hz. The frequency ($\omega_{a,b}$) and linewidth ($\gamma_{a,b}$) of each mechanical mode in the presence of the control lasers is acquired by measuring the membrane's response to an external force and fitting the data to the expected form (Methods). These provide the real and imaginary parts of the eigenvalue spectrum, and are shown as the points in Figure 2. They exhibit a sharp feature near $\Delta_{VEP}/2\pi = -15$ kHz and $P_{VEP} = 4.7$ µW, which we refer to as a virtual exceptional point (VEP) since it appears identical to an EP except that the real parts of the eigenvalues (i.e., $\omega_{a,b}$) always remain separated by ≈ 150 kHz. The smooth sheets in Fig. 2 are the result of fitting the measured $\omega_{a,b}(\Delta,P)$ and $\gamma_{a,b}(\Delta,P)$ to the real and imaginary parts of the eigenvalues of $H_{eff}$ (a rotating version of this figure is in Methods). This fit gives $g_1/2\pi$ = 2.11 Hz, $g_2/2\pi$ = 2.12 Hz, and the values of $\omega_{1,2}$ and $\gamma_{1,2}$ stated above.

One may wonder whether the VEP can be used to perform TET, since the actual mode frequencies remain well separated. To address this question, we performed a series of measurements in which one normal mode is initially excited and then $\Delta$ and $P$ are varied to execute a closed loop. Each loop is a rectangle defined by the points ($\Delta_{min}$, $P_{min}$), ($\Delta_{min}$, $P_{max}$), ($\Delta_{max}$, $P_{max}$), ($\Delta_{max}$, $P_{min}$), returning to ($\Delta_{min}$, $P_{min}$) after a total time $\tau$ = 40 ms. This value of $\tau$ is chosen to ensure adiabatic evolution during the operation; we discuss the dependence upon $\tau$ further below. Before each loop, mode 1 is driven to an energy $E \sim 10^{-19}$ J (corresponding to an amplitude $\sim 10^{-11}$ m) and mode 2 is undriven (except by thermal fluctuations, which correspond to less than 1% of the energy in mode 1 and do not qualitatively impact the results presented here). Then the drive is turned off and the control loop is implemented. After the loop we measure the energy in each of the two modes (Methods). The system is always subject to damping, so to quantify the energy transfer we define the efficiency $\varepsilon = E_f/(E_1 + E_2)$ where $E_{1,2}$ are the energies in each mode after the loop and $f$ = 2 (1) if the energy is initially stored in mode 1 (2).

Figure 3a shows $\varepsilon$ for loops in which $\Delta_{min}/2\pi$ = – 604 kHz, $P_{min}$ = 0.08 µW, $P_{max}$ = 8.3 µW, and $\Delta_{max}$ is varied. Figure 3b shows $\varepsilon$ for loops in which $\Delta_{min}/2\pi$ = – 604 kHz, $\Delta_{max}/2\pi$ = 376 kHz, $P_{min}$ = 0.08 µW, and $P_{max}$ is varied. Both figures show that when the control loop is sufficiently far from the VEP, loops enclosing the VEP result in energy transfer while loops not enclosing the VEP do not. This demonstrates the topological nature of the dynamics when the VEP is encircled.[9]

To determine whether control loops enclosing the VEP show the same nonreciprocity as for a conventional EP,[9,5,28,29] we measured $\varepsilon$ for loops encircling the VEP in either the counter-clockwise (CCW) or clockwise (CW) sense, and with the initial excitation either in mode 1 or mode 2. For each of the four possible combinations, the same loop shape was employed: $\Delta_{min}/2\pi$ = – 604 kHz, $\Delta_{max}/2\pi$ =376 kHz, $P_{min}$ = 0.08 µW, and $P_{max}$ = 8.3 µW. Figure 4 shows $\varepsilon(\tau)$ for each of these cases. In all four cases $\varepsilon \to 0$ as $\tau \to 0$, as expected for a sudden perturbation. However in the adiabatic limit we find that $\varepsilon \to 1$ when mode 1 (2) is initially excited and the loop is CCW (CW), but that $\varepsilon \to 0$ when mode 2 (1) is initially excited and the loop is CCW (CW). As in the case of a conventional EP, this nonreciprocity reflects the fact that adiabatic evolution tries to follow the topological structure of the spectrum, while the differential loss between the two modes tends to leave excitations primarily in the less-damped mode.[5,28,29]

In conclusion, we have demonstrated a simple protocol by which the standard cavity optomechanical interaction can be used to engineer exceptional points between arbitrary pairs of mechanical modes. We have used this protocol to perform nonreciprocal and topological energy transfer between modes with widely spaced frequencies. Prior to this work, topological energy transfer had relied on pairs of nearly-degenerate modes, which typically require precision fabrication and which restricts these operations to a small fraction of mode pairs. In contrast, the protocol demonstrated here can be used in any cavity optomechanical device (without the need for pre-existing nearly-degenerate modes) and can be applied to any pair of modes. This should extend topological control to a much wider range of devices, including those that are optimized for strong optomechanical coupling, incorporation into hybrid systems, and access to quantum effects.

**Acknowledgments:** We thank A. Shkarin for helpful discussions. This work is supported by AFOSR grant FA9550-15-1-0270.

**Author Contributions:** H.X., D.M. and L.J. performed the measurements and analyzed the data. H.X. and J.G.E.H. wrote the manuscript with input from all the authors. J.G.E.H. directed the research.

**Additional Information:** Supplementary information is available in the online version of the paper. Correspondence and requests for materials should be addressed to J.G.E.H.

**Competing Financial Interests:** The authors declare no competing financial interests.

**Figure Captions**

**Figure 1 | Optomechanically induced virtual exceptional point. a,** Spectrum of cavity modes and lasers. Two lasers (orange) drive a single cavity mode to generate coupling between two mechanical modes. A separate laser (green) monitors the mechanical modes' motion. **b,** A microscopic picture of the coupling induced by the two control lasers. Two processes couple the mechanical modes: in one process (hollow lines), the drive from laser 1 absorbs a phonon from mechanical mode 1 and creates a cavity photon, which then generates a phonon in mechanical mode 2 under the drive from laser 2; in the other process (solid lines) a phonon is created in mechanical mode 2 when a cavity photon is absorbed by the drive of laser 2, then the drive from laser 1 converts a phonon from mechanical mode into a cavity photon. In both processes, a phonon is transferred from mechanical mode 1 to mode 2 by the two driving lasers. This physical picture extends to the quantum case; however we note that our experiments are in the classical regime. **c, d,** The resonance frequencies (**c**) and linewidths (**d**) of the two mechanical modes. The solid points are measurements, and the sheets are the fit described in the text. In **c** the vertical axes for the two resonance frequencies have the same scale but are shifted (red and green text) to emphasize the virtual exceptional point. Each thin vertical line denotes the difference between a data point and the fit.

**Figure 2 | Topological energy transfer by encircling the virtual exceptional point. a,** Transfer efficiency $\varepsilon$ as a function of the maximum detuning $\Delta_{max}$ of the control loop. **b,** Transfer efficiency $\varepsilon$ as a function of the maximum power $P_{max}$ of the control loop. The points are the experimental results. These results show the dependence of the energy transfer on the control loop's topology, as adiabatic loops enclosing the VEP have $\varepsilon \to 1$ while adiabatic loops not enclosing the VEP have $\varepsilon \to 0$. The solid lines are numerical simulations of the dynamics governed by $H_{eff}$, using the parameters determined by the fit in Fig. 1(c,d). These simulations agree well with the data for all control loops, regardless of whether or not the loops pass close to the VEP.

**Figure 3 | Nonreciprocity of the topological energy transfer. a,** Transfer efficiency $\varepsilon$ as a function of the control loop duration $\tau$ for clockwise loops with the initial drive applied to mode 1 (red) or mode 2 (blue). **b,** The same as (**a**) but with counterclockwise loops. The points are experimental results while the solid lines are numerical simulations of the dynamics governed by $H_{\text{eff}}$, using the parameters determined by the fit in Fig. 1(c,d). Rapid loops ($\tau \to 0$) result in vanishing energy transfer $\varepsilon \to 0$ in all four cases. For adiabatic encircling, $\varepsilon$ depends on the sense of the loop and the initial condition. For counterclockwise (clockwise) loops, $\varepsilon \to 1$ as $\tau$ increases when mode 1 (2) is initially driven, and $\varepsilon \to 0$ when mode 2 (1) is initially driven.

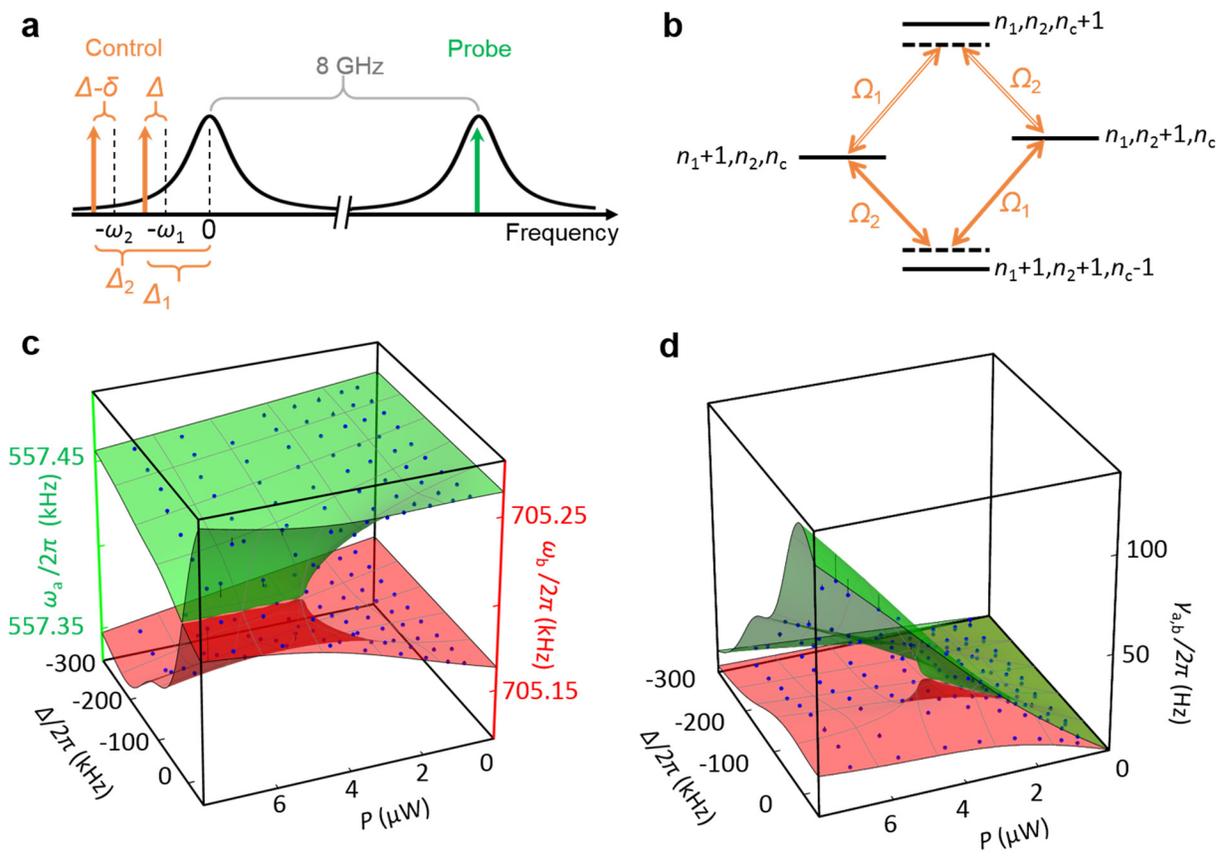

**Figure 1**

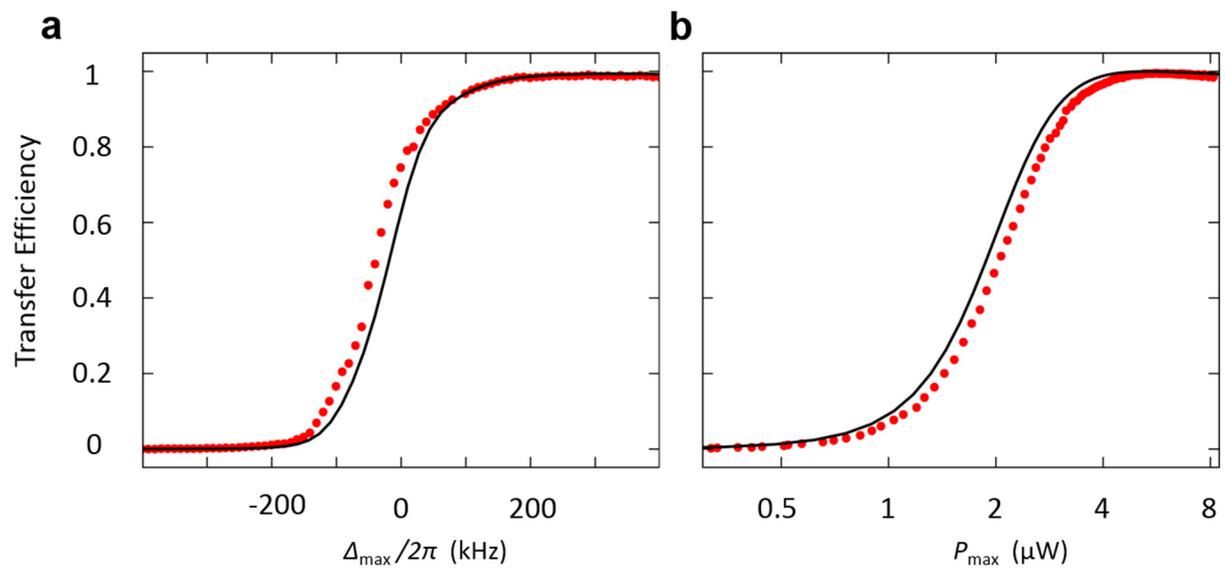

**Figure 2**

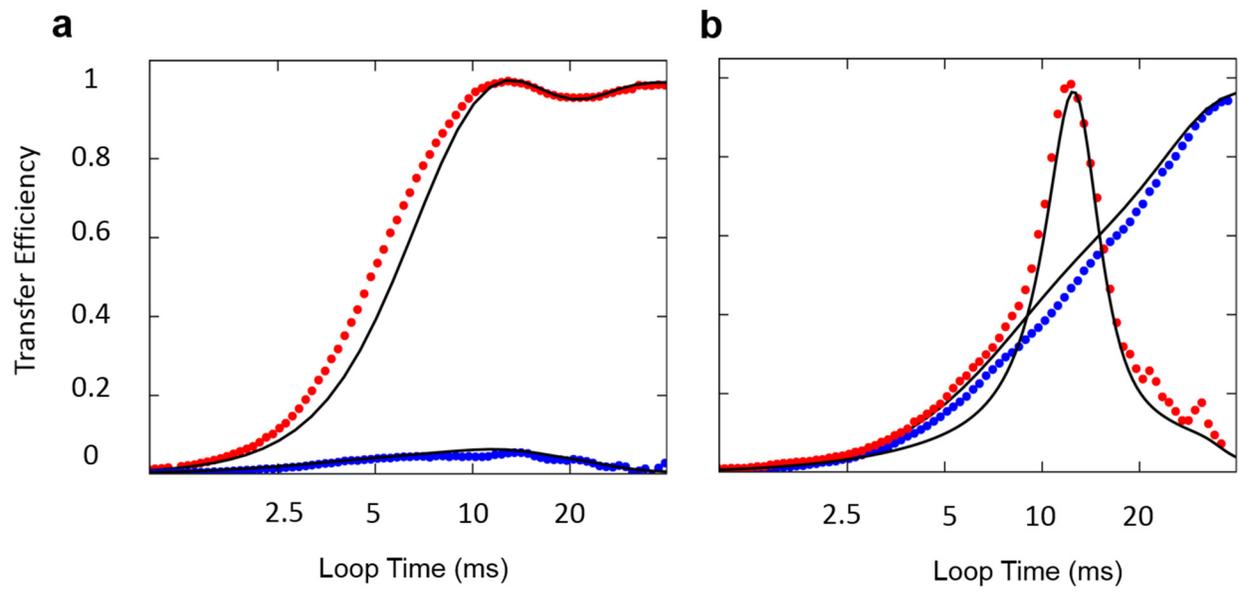

**Figure 3**

# Supplementary Information

H. Xu,[1] D. Mason,[1] L. Jiang,[1] J. G. E. Harris[1,2]

[1]Department of Physics, Yale University, New Haven, Connecticut 06511, USA
[2]Department of Applied Physics, Yale University, New Haven, Connecticut 06511, USA

## 1  Effective Hamiltonian

We consider an optomechanical system composed of a membrane and an optical cavity, driven by two control lasers. We focus on two highly non-degenerate mechanical modes, whose frequency difference is much larger than the optomechanical coupling strength (and thus a single frequency control laser would not induce an exceptional point). Here we show how to obtain the effective Hamiltonian for the mechanical modes.

Our system can be modeled by the following Hamiltonian,

$$H = \hbar(\Omega_c - i\frac{\kappa}{2})a^\dagger a + \sum_{k=1}^{2} \hbar(\omega_k - i\frac{\gamma_k}{2})c_k^\dagger c_k + \sum_{k=1}^{2} g_k(c_k + c_k^\dagger)a^\dagger a, \quad (1)$$

where $a$ and $c_k$ are the annihilation operators for photons and phonons, or the mode amplitudes in the classical regime. Considering the control lasers and neglecting the noise terms, the equations of motion are

$$\begin{aligned}
\dot{c}_1 &= -(\frac{\gamma_1}{2} + i\omega_1)c_1 - ig_1 a^\dagger a, \\
\dot{c}_2 &= -(\frac{\gamma_2}{2} + i\omega_2)c_2 - ig_2 a^\dagger a, \\
\dot{a} &= -(\frac{\kappa}{2} + i\Omega_c)a - i\sum_{k=1}^{2} g_k(c_k + c_k^\dagger)a + \sqrt{\kappa_{\text{in}}}(a_{\text{in},1}e^{-i\Omega_1 t} + a_{\text{in},2}e^{-i\Omega_2 t}),
\end{aligned} \quad (2)$$

where $a_{in,1} = \sqrt{\frac{P_k}{\hbar\Omega_k}}$ is the input field amplitude. Linearizing the above equations by $a = (\alpha + d)e^{-i\Omega_c t}$ gives

$$\begin{aligned}\dot{c}_k &= -(\frac{\gamma_k}{2} + i\omega_k)c_k - ig_k(\alpha^* d + \alpha d^\dagger),\\ \dot{d} &= -\frac{\kappa}{2}d - i\sum_{k=1}^{2} g_k(c_k + c_k^\dagger)\alpha,\end{aligned} \quad (3)$$

where $\alpha = \sum_{m=1}^{2}\sqrt{\kappa_{in}}a_{in,m}\chi(\Delta_m)e^{-i\Delta_m t} = \sum_{m=1}^{2}\alpha_m e^{-i\Delta_m t}$. The cavity's optical susceptibility is $\chi(\omega) = 1/(\kappa/2 - i\omega)$, and $\Delta_m$ is the detuning of each control laser. Integrating out the optical field, we obtain the equations of motion for the mechanical modes:

$$\begin{aligned}\dot{c}_1 &= -(\frac{\gamma_1}{2} + i\omega_1 + i\sigma_{11})c_1 - i\sigma_{12}e^{i(\Delta_1 - \Delta_2)t}c_2,\\ \dot{c}_2 &= -(\frac{\gamma_2}{2} + i\omega_2 + i\sigma_{22})c_2 - i\sigma_{21}e^{-i(\Delta_1 - \Delta_2)t}c_1,\end{aligned} \quad (4)$$

where $\sigma_{nn}$ and $\sigma_{mn}$ are defined as

$$\sigma_{nn} = \sum_{k=1}^{2}\frac{-ig_n^2 P_k}{\hbar\Omega_k}\frac{\kappa_{in}}{(\kappa/2)^2 + \Delta_k^2}\left(\frac{1}{\kappa/2 - i(\omega_n + \Delta_k)} - \frac{1}{\kappa/2 - i(\omega_n - \Delta_k)}\right),$$

$$\sigma_{mn} = \frac{-ig_1 g_2\sqrt{P_1 P_2}}{\hbar\sqrt{\Omega_1\Omega_2}}\frac{\kappa_{in}}{(\kappa/2 + i\Delta_m)(\kappa/2 - i\Delta_n)}\left(\frac{1}{\kappa/2 - i(\omega_m + \Delta_m)} - \frac{1}{\kappa/2 - i(\omega_m - \Delta_n)}\right).$$

Thus we obtain the effective Hamiltonian for the mechanical modes,

$$H_{eff} = \sum_{k=1}^{2} \hbar(\omega_k - i\frac{\gamma_k}{2} + \sigma_{kk})c_k^\dagger c_k + \sigma_{12}e^{i(\Delta_1 - \Delta_2)t}c_1^\dagger c_2 + \sigma_{21}e^{-i(\Delta_1 - \Delta_2)t}c_2^\dagger c_1. \quad (5)$$

## 2 Measurement of the mechanical spectrum

We have shown the complex spectrum (frequencies and linewidths) of the mechanical modes in Fig. 1 (c) and (d) of the main text. We obtain each experimental data point there by measuring the response of the mechanical modes to an external drive. For each pair of parameters $P$ and $\Delta$ of the control lasers, we sweep the driving frequency through the mechanical resonance frequencies, and record the response of the mechanical modes by a lock-in amplifier using the heterodyne signal of the measurement laser. We then fit the response data to complex Lorentzians, and obtain the frequencies and linewidths of the mechanical modes. An example is shown in Extended Data Fig. 1.

## 3 Measurement of energy transfer efficiency

Each measurement of the energy transfer efficiency (e.g., in Fig. 2 & 3 of the main text) begins with one of the mechanical modes driven to an energy $\sim 10^{-19}$ J (corresponding to an amplitude $\sim 10^{-11}$ m). At time $t = -10$ ms the drive is turned off. Then the control loop is carried out from $t = 0$ ms to $t = \tau$. After the control loop, the modes relax to thermal equilibrium.

In Extended Data Fig. 2 we show the energy stored in each mode as a function of time $t$ for two different control loops. In the left panel the loop is given by $\Delta_{min}$ =-604 kHz, $\Delta_{max}$ =-400 kHz, $P_{min} = 0.08$ $\mu$W, $P_{max} = 8.3$ $\mu$W, which does not enclose the EP. In the right panel the loop is given by $\Delta_{min}$ =-604 kHz, $\Delta_{max}$ =400 kHz, $P_{min} = 0.08$ $\mu$W, $P_{max} = 8.3$ $\mu$W, which does enclose the EP. For both of these loops $\tau = 40$ ms.

Since the transfer efficiency $\epsilon$ is defined in terms of $E_1$ and $E_2$ (the energies in the two modes at the end of the loop) we fit the data for $t > \tau + 10$ ms to a decaying exponential, and extrapolate this fit back to $t = \tau$ to determine $E_1$ and $E_2$.

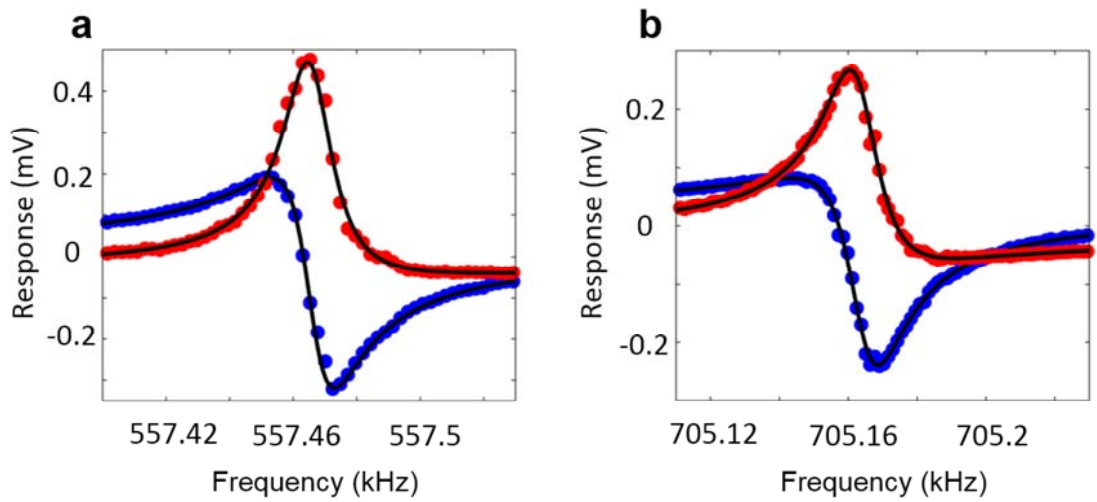

Extended Data Figure 1: Response of mechanical modes as a function of external drive frequency at $\Delta = 40$ kHz and $P = 1.3$ μW. The data points are real (blue) and imaginary parts (red) of the response of (a) the lower mode and (b) the higher mode measured by a lock-in amplifier. The solid lines are fitting to the experimental data, which give the frequencies ($557466.1 \pm 0.2$ Hz and $705164.0 \pm 0.2$ Hz) and linewidths ($19.2 \pm 0.3$ Hz and $19.5 \pm 0.5$ Hz) of the mechanical modes.

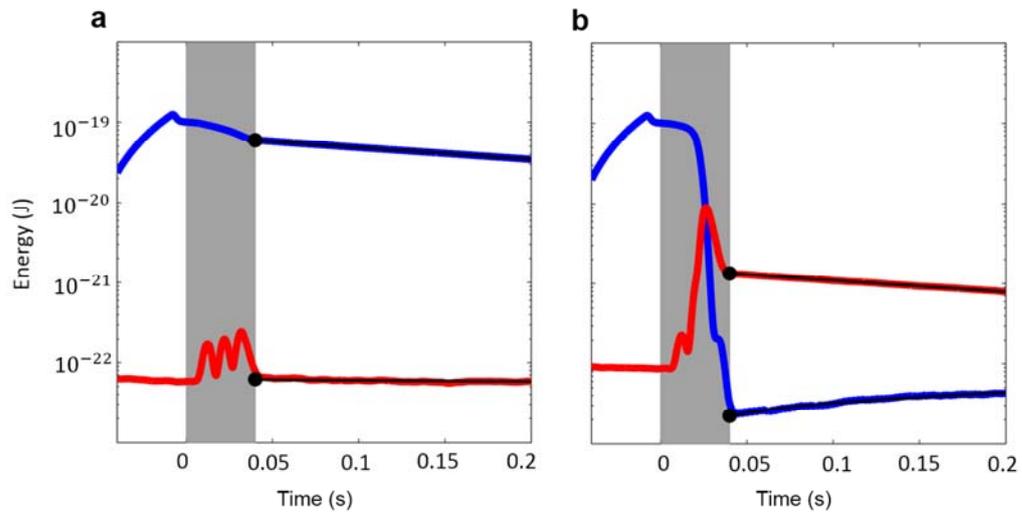

Extended Data Figure 2: Energies as a function of time $t$ for (a) a loop not enclosing the EP and (b) a loop enclosing the EP. Mode 1 (blue) is driven before $t = -10$ ms, while mode 2 (red) remains undriven. The topological operation (grey region) starts at $t = 0$ ms and ends at $t = 40$ ms, after which the modes relax to thermal equilibrium. The black lines are fits to the data for $t > 50$ ms, and the black dots denote the energies in the mechanical modes at the end of each topological operation.

5